\documentclass[reprint,author-numerical,nofootinbib]{revtex4-2}
\usepackage[utf8]{inputenc}
\usepackage{amsmath}
\usepackage{amsfonts}
\usepackage{graphicx}
\usepackage{bm}

\def\dif{{\rm d}}

\def\arsinh{{\rm arsinh}\, }
\def\arcosh{{\rm arcosh}\, }
\newcommand{\be}{\begin{equation}}
\newcommand{\ee}{\end{equation}}

\begin{document}

\preprint{AIP/123-QED}

\title[T-model field equations: the general solution]{T-model field equations: the general solution}

\author{Joan Josep Ferrando}
\altaffiliation[Also at ]{Observatori Astron\`omic, Universitat
de Val\`encia,  E-46980 Paterna, Val\`encia, Spain}
\email{joan.ferrando@uv.es.} \affiliation{ Departament d'Astronomia
i Astrof\'{\i}sica, Universitat
de Val\`encia, E-46100 Burjassot, Val\`encia, Spain.}
\author{Salvador Mengual}
\affiliation{ Departament d'Astronomia i Astrof\'{\i}sica,
Universitat de Val\`encia, E-46100 Burjassot, Val\`encia, Spain.}
%

\date{\today}

\begin{abstract}
We analyze the field equations for the perfect fluid solutions admitting a group G$_3$ of isometries acting on
orbits S$_2$ whose curvature has a gradient that is tangent to the fluid flow (T-models). We propose several methods to integrate the field equations and we present the general solution without the need to calculate any integral.
\end{abstract}

\maketitle

\section{\label{sec-intro}Introduction}

 A perfect fluid solution admitting a three-dimensional group G$_3$ of isometries acting on spacelike two-dimensional orbits S$_2$ has a metric line element that, in comoving-synchronous coordinates, takes the form \cite{kramer}: 
\begin{subequations} \label{metric-ss}
\begin{align} 
\label{metric-ss-a}
ds^2= -e^{2\nu}dt^2 + e^{2\lambda} dr^2 + Y^2 C^2 (dx^2+dy^2), \\[2mm]
%
%
\label{metric-ss-b}
\nu=\nu(r, t), \qquad \lambda=\lambda(r, t), \qquad Y=Y(r, t), \\[0mm]
%
\label{metric-ss-c}
\hspace{-4mm} C=C(x, y)\equiv\left[1+\frac{k}{4}(x^2+y^2)\right]^{-1} \! \! \! \! , \quad k=0, \pm1 ,
\end{align} 
\end{subequations}
where the value of $k$ distinguishes the plane, spherical and hyperbolic symmetries. Moreover, the metric functions (\ref{metric-ss-b}) are submitted to two second order differential equations as a consequence of the perfect fluid conditions, namely, $G_r^r = G_x^x$ and $G_r^t=0$.

When the curvature of the orbits S$_2$ has a gradient that is tangent to the fluid flow, that is, when $Y= Y(t)$, one says that the solution is a {\em T-model}. The notions of {\em T-region} and {\em R-region} were introduced for the spherically symmetric case by Novikov \cite{Novikov-62} who also discussed the solutions that are T-regions globally \cite{[{}][{ [English translation: 1964 {\em Sov. Astr. A. J.} {\bf 7} 587]}] Novikov-63} (see also \cite{[{}][{ [English translation: 2001 {\em Gen. Relativ. Gravit.} {\bf 33} 2259]}] Novikov-64}). 

Ruban \cite{Ruban} showed that the spherically symmetric perfect fluid T-models have geodesic motion (see also \cite{Krasinski}), a result that can be extended to the plane and hyperbolic symmetries (see, for example, \cite{Krasinski-Plebanski}). Thus, the T-models are the perfect fluid solutions whose metric has the form (\ref{metric-ss}), with $\nu=\nu(t)$ and $Y=Y(t)$. 

The spherical dust T-model was first considered by Datt \cite{[{}][{ [English translation: 1999 {\em Gen. Relativ. Gravit.} {\bf 31} 1615]}] Datt}, and the dust solution with cosmological constant was widely analyzed later by Ruban, who showed that this solution has no Newtonian analog \cite{[{}][{ [English translation: 1968 {\it Sov. Phys. JETP Lett.} {\bf 8} 414] [Reprinted: 2001 {\em Gen. Relativ. Gravit.} {\bf 33} 363]}] Ruban-68, [{}][{ [English translation: 1969 {\it Sov. Phys. JETP} {\bf 29} 1027] [Reprinted: 2001 {\em Gen. Relativ. Gravit.} {\bf 33} 375]}] Ruban}.  The perfect fluid T-models with a nonconstant pressure were examined by Korkina and Martinenko \cite{Korkina} and Ruban \cite{[{}][{ [English translation: 1983 {\it Sov. Phys. JETP} {\bf 58} 463]}]Ruban-83}, while Herlt \cite{Herlt} proposed an algorithm to obtain new solutions in this family (see also \cite{kramer}). The spatially homogenous limit of the T-models ($\lambda= \lambda(t)$) were considered by Kompanneets and Chernov \cite{[{}][{ [English translation: 1965 {\em Sov. Phys. JETP} {\bf 20} 1303]}]Kompa} and were later studied by Kantowski and Sachs \cite{K-S} for a dust source (see \cite{Krasinski} for more references). 

Understanding the physical meaning of the T-models is still an open problem. To take a small step towards this goal we have recently proposed a thermodynamic interpretation of these solutions \cite{FM-Tmodels}. In this reference we have obtained the thermodynamic schemes associated with a specific T-model, and we have determined the solutions that can model a generic ideal gas. On the other hand, we have generalized and analyzed from a thermodynamic point of view the McVittie-Wiltshire-Herlt solution. This T-model can be obtained by applying the Herlt algorithm to the homogeneous T-model presented by McVittie and Wiltshire \cite{McVittie}. 

In any case, there are very few T-models for which we know the explicit analytic expression of the metric functions, and it would be suitable to know more solutions for a better understanding of the physical and geometric properties of the T-models. 

In this paper we analyze the field equations for the T-models, we revisit the Herlt integration algorithm, and we propose new ones that provide the general solution without making any indefinite integral. 


In Sec. \ref{sec-Tmodels}, we revisit the perfect fluid field equations for the T-models and we point out that the space of solutions is controlled by two arbitrary real functions, one depending on the time-coordinate $t$ and another depending on the spatial coordinate $r$. We also give the expression of the energy density $\rho$, the pressure $p$ and the expansion $\theta$ in terms of the metric functions. In Sec. \ref{subsec-termo-Tmodels} we comment about our results on the thermodynamics of a T-model solution \cite{FM-Tmodels}, and we give the function of state $c_s = \chi(\rho,p)$ that provides the square of the speed of sound in terms of the energy density and the pressure. In Sec. \ref{subsec-ideal} we summarize briefly the ideal T-models analyzed in \cite{FM-Tmodels}, which are compatible with the ideal gas equation of state. Some new solutions are presented in Sec. \ref{subsec-v=1-knot=0}.

In Sec. \ref{sec-integracio-1} we analyze the Herlt algorithm and we show that implementing the algorithm to calculate the solution requires the realization of two indefinite integrals. We also propose another integration algorithm to obtain the solution by quadratures. As in the Herlt algorithm, obtaining an indefinite integral detects a homogeneous solution, and obtaining another integral determines a nonhomogeneous T-model.

In Sec. \ref{sec-solucio-k=0} we obtain the general solution in the case of plane symmetry ($k=0$). We determine the metric line element, and the hydrodynamic quantities $\rho$, $p$ and $\chi(\rho,p)$, in terms of two arbitrary real functions $\{\varphi(t), Q(r)\}$. Moreover, we recover previously known solutions, and we obtain new ones.  

In Sec. \ref{sec-solucio-knot=0} we redefine the metric functions in such a way that the field equations become algebraic in one of the unknown functions. This fact allows us to obtain the general solution for spherical and hyperbolic symmetries ($k\not=0$). We provide two different algorithms to determine the solution in terms of an arbitrary function $Q(r)$ of the spatial coordinate $r$ and an arbitrary function of  $t$.

Finally, in Sec. \ref{sec-remarks} we comment on the results obtained here and we explain how our results also apply for the homogeneous T-models and for their generalization without symmetries.


\section{Field equations for the T-models}
\label{sec-Tmodels}

In \cite{FM-Tmodels} we have shown that the field equations for the T-models can be written as a second order differential equation which is linear for a specific choice of the metric functions. Indeed, if we make $e^{\lambda} = \omega(t,r)>0$, $e^{-2 \nu} = v(t)>0$ and $Y^2 = \varphi(t)>0$, then the metric line element (\ref{metric-ss}) becomes
\begin{equation} \label{metric-T-1}
ds^2= -\frac{1}{v(t)}dt^2 + \omega^2(t,r) dr^2 + \varphi(t) C^2 (dx^2+dy^2),
\end{equation}
where $C$ is given in (\ref{metric-ss-c}). Now, the perfect fluid field equation $G_r^t=0$ identically holds, and $G_r^r = G_x^x$ holds if, and only if, the metric functions $v(t)$, $\omega(t,r)$ and $\varphi(t)$ meet the differential equation
\be \label{eq-T-1}
2 v \varphi \, \ddot{\omega} + (\dot{v} \varphi + v \dot{\varphi})\, \dot{\omega} - (v \ddot{\varphi} + \frac12 \dot{v} \dot{\varphi} + 2k)  \,\omega = 0 \, ,
\ee
where a dot denotes derivative with respect to the time coordinate $t$.

The unit velocity of the fluid $u = \sqrt{v} \, \partial_t$ is geodesic and its expansion is
\begin{equation} \label{expansion-T-1}
\theta = \sqrt{v} \left(\frac{\dot{\varphi}}{\varphi} + \frac{ \dot{\omega}}{\omega}\right)  = \sqrt{v} \, \partial_t [\ln(\varphi \omega)] \,  .
\end{equation}
And the {\em pressure} $p$ and the {\em energy density} $\rho$ are then given by
\begin{eqnarray} \label{pressure-T-1}
p =  v \left[\frac14 \frac{\dot{\varphi}^2}{\varphi^2} - \frac{\ddot{\varphi}}{\varphi} - \frac12 \frac{\dot{\varphi}}{\varphi}  \frac{ \dot{v}}{v}\right] - \frac{k}{\varphi}   \,  ,  \\[1mm]
\rho =  v \left[\frac14 \frac{\dot{\varphi}^2}{\varphi^2} +  \frac{\dot{\varphi}}{\varphi}  \frac{\dot{\omega}}{\omega}\right] + \frac{k}{\varphi}   \,  .
\label{density-T-1}
\end{eqnarray}

The known T-models have usually been obtained by considering the functions $v(t)$, $\omega(t,r)$ and $Y(t)$ as unknown metric functions. The field equations are linear in the functions $v(t)$ and $\omega(t,r)$, and this fact plays an important role in the integration process. Note that our choice of the metric function $\varphi = Y^2$ as an unknown of the field equations, leads us to Eq. (\ref{eq-T-1}), which is also a linear equation for $\varphi$. Thus, this equation is linear for the three involved metric functions, a significant quality that will help us in our approach.

The spatially homogeneous limit of the T-models are the Kompanneets-Chernov-Kantowski-Sachs (KCKS) metrics \cite{Kompa, K-S}. In fact, from the expressions of the expansion (\ref{expansion-T-1}) and the energy density (\ref{density-T-1}), we obtain the following four equivalent conditions that characterize these solutions:
\begin{itemize}
\item[(i)]
The metric function $\omega(t,r)$ factorizes. And then, one can take the coordinate $r$ so that $\omega' = \partial_r \omega = 0$, that is, $\omega= \omega(t)$.
\item[(ii)]
The spacetime is spatially homogeneous. And then, it admits a group G$_4$ of isometries acting on orbits S$_3$.
\item[(iii)]
The energy density is homogeneous, $\rho = \rho(t)$. And then, the fluid has a barotropic evolution.
\item[(iv)]
The fluid expansion is homogeneous, $\theta = \theta(t)$.
\end{itemize}

Note that (\ref{eq-T-1}) is a homogeneous linear second order differential equation for the function $\omega(t,r)$ when $v(t)$ and $\varphi(t)$ are given. Then, we can choose the coordinate $r$ so that \cite{FM-Tmodels}
\be \label{w-w1-w2}
\omega(t,r) =  \omega_1(t)  + \omega_2(t)  \, Q(r) \, ,
\ee
where $Q(r)$ is an arbitrary real function, and $\omega_i(t)$ being two particular solutions to the Eq. (\ref{eq-T-1}). 

The spacetime metric does not change with a redefinition of the time coordinate, $t = t(T)$. Every choice of $t$ can be realized by imposing a constraint on the time-dependent functions $v(t)$, $\varphi(t)$ and $\omega_i(t)$. This coordinate condition, and Eq. (\ref{eq-T-1}) imposed on each of the functions $\omega_i$, constitute a set of three constraints for the four metric functions $\{\varphi(t), \omega_i(t), v(t)\}$. Consequently, the space of solutions depends on an arbitrary real function depending on time, and another real function, $Q(r)$, depending on $r$.

It is quite usual in literature (see, for example, \cite{kramer, Krasinski}) to choose the time coordinate such that $t= Y = \sqrt{\varphi}$. Then, the functions $\omega_i(t)$ are determined by Eq. (\ref{eq-T-1}) if we give the function $v(t)$. In this case, the space of solutions is controlled by the functions $\{v(t), Q(r)\}$. 

Alternatively, we can give as input one of the functions $\omega_i$, say $\omega_2$, and then Eq. (\ref{eq-T-1}) becomes a first order linear differential equation for the function $v(t)$; once this equation is solved, we can proceed to determine $\omega_1$ by once again using (\ref{eq-T-1}) with the $v(t)$ previously obtained. This procedure by Herlt \cite{Herlt} shows that the field equation can be solved by quadratures, and the space of solutions is controlled by the functions $\{\omega_2(t), Q(r)\}$. 

In our recent thermodynamic approach to the T-models \cite{FM-Tmodels} we have taken as time coordinate the proper time $\tau$ of the Lagrangian observer associated with the fluid. This means that $v(\tau)=1$, and then, for every choice of the function $\varphi(\tau)$, Eq. (\ref{eq-T-1}) determines two particular solutions $\omega_i(\tau)$. Thus, with this choice, the space of solutions is controlled by the functions $\{\varphi(\tau), Q(r)\}$.


\subsection{Thermodynamics of the T-models}
\label{subsec-termo-Tmodels}


When does a perfect fluid solution represent the evolution in local thermal equilibrium of a realistic perfect fluid? What are its thermodynamic properties? A precise theoretical framework in which to answer these questions has been developed in \cite{Coll-Ferrando-termo, CFS-LTE, CFS-CC}, and it has been applied to analyze some families of perfect fluid solutions  \cite{CF-Stephani, CFS-CIG, CFS-RSS, CFS-PSS}. The indicatrix function of the local thermal equilibrium, $\chi = u(p)/u(\rho)$, plays a central role in our procedure (for a function $f(x^{\alpha})$, $u(f) = u^{\alpha} \partial_{\alpha} f$). When $\chi$ is a function of state, $\chi =\chi (\rho ,p)$, it physically represents the square of the speed of sound in the fluid, $\chi (\rho ,p) \equiv  c^2_{s}$ \cite{CFS-RSS}. 

Recently we have carried out this thermodynamic approach to the T-models \cite{FM-Tmodels}, and we have obtained the general expression of the indicatrix function when $v(\tau)=1$. Without this choice of the time coordinate, a similar calculation leads to
\be  \label{chi-Tmodels}
c_s^2  = \frac{u(p)}{u(\rho)} = \chi(\rho,p)  \equiv \frac{1}{{\cal A}(p) \rho^2 + {\cal B}(p) \rho + {\cal C}(p)} \, ,
\ee
where ${\cal A}$, ${\cal B}$ and ${\cal C}$ are the functions of $t$ (and then of $p$ trough (\ref{pressure-T-1})) given by
\begin{subequations} \label{ABCcal}
\begin{align}
\hspace{-2mm} {\cal A}(p) \equiv  -\frac{1}{v \, b \, \dot{p}}  , \ \ {\cal B}(p) \equiv  {\cal A} \, (p + q)  ,  \ \ {\cal C}(p) \equiv  {\cal A} \, p \,q  \label{ABCcal-a} \,  , \\ 
b(t) \equiv \frac{\dot{\varphi}}{\varphi} \, , \qquad q \equiv \frac34 v \, b^2 - \frac{k}{\varphi} \, . \qquad  \qquad 
\label{b-q}
\end{align} 
\end{subequations}

The indicatrix function collects all the thermodynamic properties that can be established using only hydrodynamic variables $\{u, \rho, p\}$, that is, those that are determined by and, in turn, constraint the gravitational field \cite{CFS-LTE, CFS-CC, CFS-RSS}. A specific thermodynamic perfect fluid solution can be furnished with a family (depending on two real functions) of thermodynamic schemes that complete the thermodynamic properties and afford different interpretations of the solution \cite{CFS-LTE}. Each thermodynamic scheme provides a set of thermodynamic quantities, $\{n, s, \Theta, \epsilon\}$ (mass density, specific entropy, temperature, specific internal energy), constrained by the common thermodynamic laws. In \cite{FM-Tmodels} we have also obtained all the thermodynamic schemes that can be associated with a given T-model.


\subsection{The ideal T-models}
\label{subsec-ideal}


The determination and subsequent study of the solutions with a specific thermodynamic behavior is a subject to be considered in the thermodynamic analyses of a family of perfect fluid solutions. 

In \cite{FM-Tmodels} we have obtained the T-models that are compatible with the equation of state of a generic ideal gas, $p= \tilde{k} n \Theta$, that is, those compatible with the ideal sonic condition $\chi = \chi(\pi) \not=1$, $\pi \equiv p/\rho$ \cite{CFS-LTE}. These solutions have, necessarily, plane symmetry $k=0$ and, with the choice $v(\tau)=1$, the metric functions $\varphi(\tau)$ and $\omega(\tau)$ take the expressions
\begin{subequations}  \label{phi-alpha-ideal}
\begin{align} 
\label{phi-alpha-ideal-a}
\varphi(\tau) = |\tau|^{\frac{4}{3 \gamma}}   ,  \qquad \omega(\tau) = \sqrt{\varphi(\tau)} [\alpha(\tau) + Q(r)] \, , \\[0mm]
\alpha(\tau) = \begin{cases} |\tau|^{1- \frac{2}{\gamma}}  , \, \qquad {\rm if} \ \ \, \gamma \not= 2  \cr 
  \ln |\tau|  , \quad \quad \ \ \,  {\rm if} \quad  \gamma = 2 \, .  \label{phi-alpha-ideal-b} 
 \end{cases}
\end{align} 
\end{subequations}
It is worth remarking that the above {\em ideal T-models} fulfill the macroscopic necessary constraints for physical reality (energy conditions, compressibility conditions, positivity of some thermodynamic quantities) in wide space-time domains \cite{FM-Tmodels}.


\subsection{Some new solutions with $k \not=0$}
\label{subsec-v=1-knot=0}


When $v(\tau)=1$, the field equation (\ref{eq-T-1}) becomes
\be \label{eq-T-v=1}
2 \varphi \, \ddot{\omega} +  \dot{\varphi}\, \dot{\omega} - (\ddot{\varphi} + 2k)  \,\omega = 0 \, .
\ee
If we consider the case $\gamma =4/3$ in the family of the ideal T-models quoted in the above subsection, we have $\ddot{\varphi} = 0$ and $2 \varphi \, \ddot{\omega} +  \dot{\varphi}\, \dot{\omega} = 0$. We can extend this solution to nonplane symmetry, $k = \pm 1$, by imposing on $\omega(\tau)$ this last equation and by considering $\varphi(\tau)$ such that $\ddot{\varphi} + 2k =0$. Then, we can introduce the change of time $\tau = \kappa \, t$, so that the solution to this equation can be expressed as
\be
\varphi(t) = \kappa^2(\varepsilon - k t^2) \, ,  \qquad \kappa >0 \, , \label{phi-knot0}
\ee
where $\varepsilon = \pm 1$ if $k=-1$, and $\varepsilon = + 1$ if $k=1$. Moreover,  $\omega_1(t)=1 $ is a particular solution to the Eq. (\ref{eq-T-v=1}), and another one is:
\begin{eqnarray}
{\rm If} \ \  k= +1, \ \ \ \qquad  \qquad \qquad \omega_2(t) = \arcsin t \,  , \, \qquad \qquad  \label{omega-k=1}  \\  
{\rm If} \ \  k= -1, \quad \varepsilon = +1: \qquad \, \omega_2(t) = \arsinh t \, ,  \qquad  \qquad    \label{omega-k=-1+}\\
{\rm If} \ \  k= -1, \quad \varepsilon = -1: \qquad  \omega_2(t) = \arcosh t \, .  \qquad  \qquad     \,  \label{omega-k=-1-}
\end{eqnarray}
Then, the pressure and the energy density take the expressions
\begin{eqnarray} \label{pressure-T-v=1}
p =  \frac{k \, \varepsilon}{\kappa^2 (\varepsilon - k t^2)^2}    \,  ,  \\[1mm]
\rho =  p \left[ 1 - \frac{ 2 t \sqrt{\varepsilon - k t^2} Q }{\varepsilon [1 + \omega_2(t) Q]} \right]   \,  .
\label{density-T-v=1}
\end{eqnarray}
And, from the expression (\ref{chi-Tmodels}, \ref{ABCcal}) of the indicatrix function of a T-model, we obtain that the square of the speed of sound is given by
\be  \label{chi-Tmodels-v=1}
\chi(\rho,p)  = \frac{8 p^2 (1 - \frac{\varepsilon}{\kappa} \sqrt{\frac{k \varepsilon}{p}})}{\rho^2 + 4 \rho \, p (1 - \frac{\varepsilon}{\kappa} \sqrt{\frac{k \varepsilon}{p}})+ p^2 (3 - 4 \frac{\varepsilon}{\kappa} \sqrt{\frac{k \varepsilon}{p}})}  ,
\ee
\begin{figure*}
\centerline{
\parbox[c]{0.51\textwidth}{\includegraphics[width=0.49\textwidth]{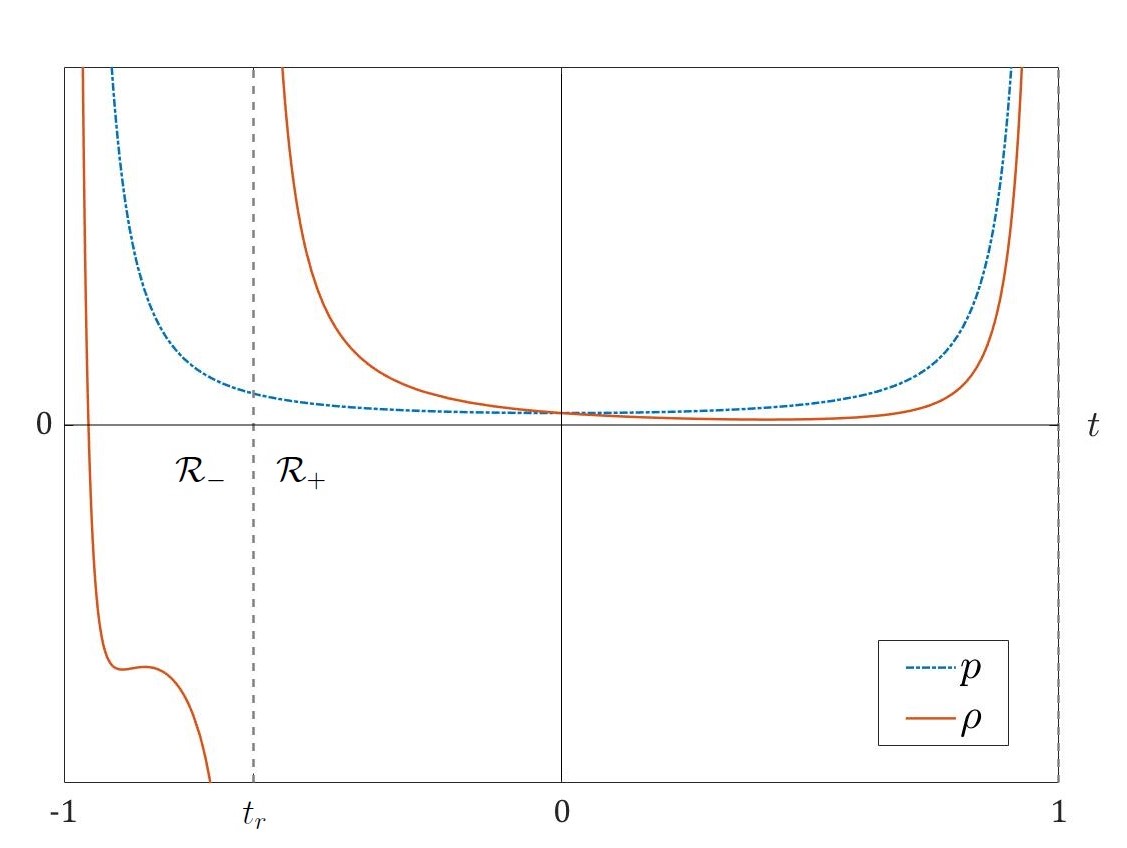}}
\parbox[c]{0.53\textwidth}{\includegraphics[width=0.49\textwidth]{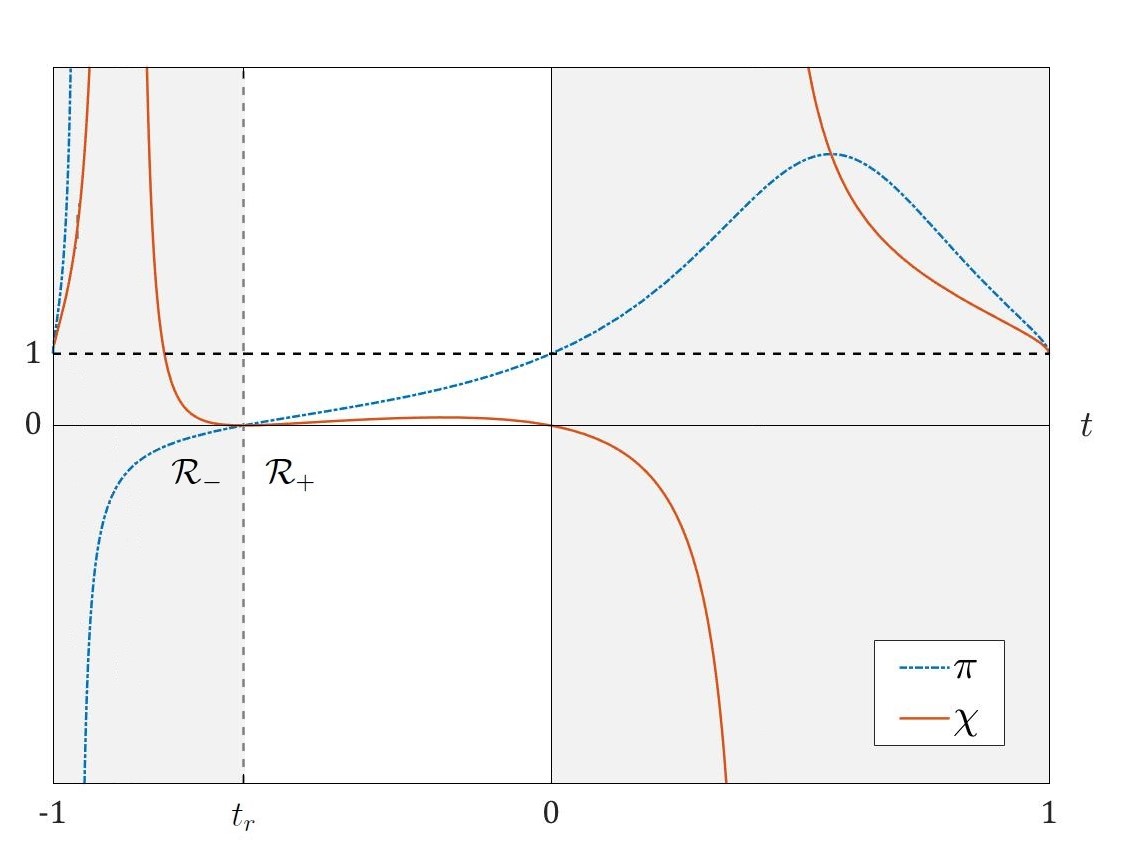}}}
\caption{This figure shows the behavior of the hydrodynamic variables of the spherically symmetric T-model defined by the functions (\ref{phi-knot0}, \ref{omega-k=1}). We have considered the case $Q_r >0$. The case $Q_r<0$ follows by changing $t$ by $-t$. On the left, we have plotted the dependence on time of the energy density $\rho$ and the pressure $p$ for $Q_r = 1.5$. Note that $\rho$ is positive in the whole region ${\cal R}_+$. On the right, we have plotted the quotient $\pi=p/\rho$, and the square of the speed of sound  $\chi = u(p)/u(\rho)$. Notice that the energy conditions ($0 < \pi < 1$), and the causal sonic condition ($0 < \chi < 1$) only hold in the subregion $]t_r, 0[$ of the region ${\cal R}_+$ (unshaded interval).}
\label{Fig-1}
\end{figure*}

The analysis of these solutions, which we do not describe in detail here, shows that their good physical behavior is constrained to limited spacetime domains:
\begin{itemize}
\item[(i)]
The spherically symmetric case, $\omega_2(t) = \arcsin t$, $t \in[-1, 1]$, leads to a positive pressure everywhere. The metric has a curvature singularity at $t= t_r \equiv - \sin [1/Q(r)]$ that disconnects two spacetime regions ${\cal R}_-$  ($t< t_r$) and ${\cal R}_+$ ($t> t_r$). In the spacetime domain where $Q(r)>0$ (respectively, $Q(r)<0$) the energy density is positive in the region ${\cal R}_+$, (respectively, ${\cal R}_-$),  as the left diagram in Fig. \ref{Fig-1} shows. Moreover, there is always a spacetime domain in which the macroscopic conditions for physical reality hold (see right diagram in Fig. \ref{Fig-1}). 
\item[(ii)]
The case $\omega_2(t) = \arsinh t$ leads to a negative pressure everywhere. Moreover, whatever the values of $Q(r)$ the energy conditions and the compressibility conditions do not hold simultaneously for any value of $t$.
\item[(iii)]
The case $\omega_2(t) = \arcosh t$ leads to a positive pressure everywhere. Moreover, in the domain where $Q(r)<0$ the energy conditions and the compressibility conditions hold simultaneously in an interval of time $t \in [1, t_1[$, $t_1 < t_r \equiv \cosh [1/Q(r)]$. The metric has a curvature singularity at $t=t_r$.

\end{itemize}


\section{Integration algorithms}
\label{sec-integracio-1}


\subsection{The Herlt algorithm}
\label{subsec-Herlt}

The field equation (\ref{eq-T-1}) is a first order linear differential equation for the metric function $v(t)$ that can be written as
\be \label{eq-Herlt}
A \dot{v} + 2 \dot{A} v - 2 k \omega = 0 \, , \qquad A \equiv \varphi \dot{\omega} - \frac12 \dot{\varphi} \omega \, .
\ee

Herlt \cite{Herlt} proposed an integration algorithm based on this fact. He considers the spherically symmetric case, chooses the time coordinate as $t=Y = \sqrt{\varphi}$ and he establishes the following steps (that we report with our notation):
\begin{itemize}
\item[h1]
Choose an arbitrary function $\omega_2(t)$.
\item[h2]
Set Eq. (\ref{eq-Herlt}) for $v(t)$ by taking  $\omega= \omega_2(t)$ and $\varphi = t^2$, and obtain the general solution $v(t)$. 
\item[h3]
Set Eq. (\ref{eq-T-1}) for the function $\omega(t)$, by taking $\varphi = t^2$ and $v(t)$ the function obtained in step 2, and obtain a particular solution $\omega_1(t)$. 
\end{itemize}

Herlt \cite{Herlt} remarked that steps 1 and 2 of his algorithm determine a homogeneous KCKS T-model, and step 3 completes a nonhomogeneous solution. He applies this algorithm to obtain a nonhomogeneous T-model from the homogeneous one presented by McVittie and Wiltshire \cite{McVittie} in which $\omega_2 = t^n$. 

It is worth remarking that the Herlt algorithm provides the solution by quadratures. Indeed, two integrals determine the solution $v(t)$ of the nonhomogeneous linear first order differential equation (\ref{eq-Herlt}). And, if we know a particular solution $\omega_2(t)$ of the homogeneous linear second order differential equation (\ref{eq-T-1}), then we can obtain another solution $\omega_1(t)$ with two indefinite integrals.

Now we revisit the Herlt algorithm and we show that: (i) it can be generalized to the plane and hyperbolic symmetries, (ii) it can be implemented without any specific choice of the time coordinate $t$, and (iii) it is only necessary to obtain two indefinite integrals to get the solution.

Let's take two arbitrary functions $\{\varphi(t), \omega_2(t)\}$, which fix the time coordinate $t$ and a solution of the field equations (for every $Q(r)$). Note that the general solution of the homogeneous equation associated with equation (\ref{eq-Herlt}) can be obtained without any integral, and it is $v_0(t) = C A^{-2}$. Then, the function $C(t) \equiv v(t) A^2(t)$ fulfills the equation $\dot{C} = 2 k \omega_2 A$, and therefore:
\be
C(t) = K_0 + 2k \! \int \! \omega_2(t) A(t) \dif t \, , \quad  K_0 = constant \, .
\ee
Consequently, we have obtained $v(t)$ by performing a single quadrature. Furthermore, $\omega_1 = \omega_2 \zeta$ is an independent solution to the homogenous linear equation (\ref{eq-T-1}) if, and only if, function $\zeta(t)$ is nonconstant and fulfills the second order differential equation
\be
2 \frac{\ddot{\zeta}}{\dot{\zeta}} + 4 \frac{\dot{\omega}_2}{\omega_2} + \frac{(\varphi v)^{\cdot}}{\varphi v} = 0 \, .
\ee
This equation is equivalent to $\dot{\zeta}^2 \omega_2^4 \varphi v = K_1^2$, where $K_1$ is a nonvanishing constant. Consequently, we obtain $\zeta(t)$ (and then $\omega_1(t)$) by taking a single quadrature:
\be
\zeta(t) = \pm K_1 \! \int \! \frac{\dif t}{\omega_2^2(t) \sqrt{\varphi(t) v(t)}} \, .
\ee
Note that, being $Q(r)$ an arbitrary function, $\zeta(t)$ can be redefined by an arbitrary constant. Following this line of reasoning we arrive to the following performance of the Herlt algorithm: 
\begin{itemize}
\item[H1]
Choose two arbitrary functions $\{\varphi(t), \omega_2(t)\}$, and obtain the function $A(t) \equiv \varphi(t) \dot{\omega}_2(t) - \frac12  \dot{\varphi}(t) \omega_2(t)$. 
\item[H2]
Determine the indefinite integral
\be
H(t)= \! \! \int \! \! \omega_2(t) A(t)\, \dif t \, , 
\ee
and obtain the metric function
\be
v(t) = \frac{1}{A^2(t)} [K_0 + 2k H(t)]  \, .
\ee
\item[H3]
Determine the indefinite integral
\be
\zeta(t) =  \!  \int \!  \frac{A(t) \dif t}{\omega_2^2(t) \sqrt{\varphi(t)}\sqrt{K_0 + 2 k H(t)}}  \, ,
\ee
and obtain the metric function
\begin{equation}
\omega(t,r) = \omega_2(t)[ \zeta(t) + Q(r)] \, ,
\end{equation}
\end{itemize}
where $Q(r)$ is an arbitrary real function. Then, the metric functions $\{\varphi(t), \omega(t,r), v(t)\}$ define a T-model (\ref{metric-T-1}) that is a solution of the field equation (\ref{eq-T-1}). 

Note that this algorithm allows us to solve the field equation by quadratures. Nevertheless, only in few cases the indefinite integrals can be calculated to obtain an explicit expression of the solution. For example, Herlt \cite{Herlt} considered $\varphi = t^2$ and $\omega_2= t^n$ in the spherically symmetric case $k=1$. The second step in the above algorithm gives
\be
v(t) = \frac{1}{n^2 -1} + C_0 \, t^{-2(n+1)} , 
\ee
which corresponds to the homogeneous solution by McVittie and Wiltshire \cite{McVittie}. The third step, which determines the function $\zeta(t)$, cannot be explicitly achieved for an arbitrary value of the constant $C_0$. When $C_0=0$ we obtain an inhomogeneous solution with $\zeta(t)= t^{-2n}$. It is worth remarking that this McVittie-Wiltshire-Herlt T-model, and its generalizations to $k=0$ and $k=-1$, do not fulfill the macroscopic necessary constraints for physical reality \cite{FM-Tmodels}.

From now on, we look in this paper for other algorithms, which are alternative to the Herlt one, that will allow us to obtain new T-model solutions.


\subsection{Field equations for the variables $(\varphi, \alpha, v)$}
\label{subsec-alpha}

Let's consider the function $\alpha(t,r)$ defined by the condition $\omega = \alpha \sqrt{\varphi}$. Then, in terms of the metric functions $\{\varphi, \alpha, v\}$, the metric tensor (\ref{metric-T-1}) becomes
\begin{equation} \label{metric-T-2}
ds^2= -\frac{1}{v(t)}dt^2 +  \varphi(t)[\alpha^2(t,r) dr^2 + C^2 (dx^2+dy^2)],
\end{equation}
where $C$ is given in (\ref{metric-ss-c}). Moreover, the field equation (\ref{eq-T-1}) takes the expression
\be \label{eq-T-2}
2 v \varphi \, \ddot{\alpha} + (\dot{v} \varphi + 3 v \dot{\varphi})\, \dot{\alpha} - 2k \alpha = 0 \, .
\ee

On the other hand, the pressure keeps the expression (\ref{pressure-T-1}), and the expansion (\ref{expansion-T-1}) and the energy density (\ref{density-T-1}) become
\begin{eqnarray} 
\label{expansion-T-2}
\theta = \sqrt{v} \left(\frac32 \frac{\dot{\varphi}}{\varphi} + \frac{ \dot{\alpha}}{\alpha}\right)  = \sqrt{v} \, \partial_t [\ln(\varphi^{3/2} \alpha)] \,  , 
\\[1mm]
\rho =  v \left[\frac34 \frac{\dot{\varphi}^2}{\varphi^2} +  \frac{\dot{\varphi}}{\varphi}  \frac{\dot{\alpha}}{\alpha}\right] + \frac{k}{\varphi}   \,  .
\label{density-T-2}
\end{eqnarray}

Note that (\ref{eq-T-2}) is a nonhomogeneous linear first order differential equation for both $v(t)$ and $\varphi(t)$, and a homogeneous linear second order differential equation for the function $\alpha(t,r)$. We have then:
\be \label{alpha-1-2}
\alpha(t,r) =  \alpha_1(t) + \alpha_2(t)  \, Q(r)\, ,
\ee
where $Q(r)$ is an arbitrary real function, and $\alpha_i(t)$ being two particular solutions to the Eq. (\ref{eq-T-2}). Thus, the four metric functions $\{\varphi(t), \alpha_i(t), v(t)\}$ are submitted to two differential equations and a constraint that fixes the time coordinate. Consequently, the space of solutions depends on an arbitrary real function depending on time, and another real function, $Q(r)$, depending on $r$.


\subsection{The modified Herlt algorithm}
\label{subsec-Herlt-A}

Given two arbitrary functions $\{\varphi(t), \alpha_2(t)\}$, the general solution of the homogeneous equation associated with Eq. (\ref{eq-T-2}) for $v(t)$ is $v_0(t) = D \varphi^{-3} \dot{\alpha}_2^{-2}$, $D$ being a constant. Then, the function $D(t) \equiv v(t) \varphi^3(t) \dot{\alpha_2}^2(t)$ fulfills equation $\dot{D} = 2 k \alpha_2 \dot{\alpha}_2 \varphi^2$, and, consequently, we can obtain $v(t)$ by performing a single quadrature. 

Furthermore, $\alpha_1 = \alpha_2 \zeta$ is an independent solution to the homogenous linear equation (\ref{eq-T-2}) if, and only if, the function $\zeta(t)$ is nonconstant and fulfills the same second order differential equation than in the Herlt algorithm, which now leads to $\dot{\zeta}^2 \alpha_2^4 \varphi^3 v = K_1^2$. Consequently, we obtain $\zeta(t)$ (and then $\alpha_1(t)$) by taking a single quadrature.

The factor $\dot{\alpha}_2$ appears in the two functions that we must integrate to obtain the solution. Thus, it is now suitable to choose the time coordinate $t$ such that $\alpha_2(t)=t$. Then, following a similar line of reasoning to that in Sec. \ref{subsec-Herlt} we arrive to the following integration algorithm: 
\begin{itemize}
\item[A1]
Choose two arbitrary real functions $\{\varphi(t), Q(r)\}$. 
\item[A2]
Determine the indefinite integral
\be
D(t) =  \! \int \! t \, \varphi^2(t) \dif t \, , 
\ee
and obtain the metric function
\be \label{A-v}
v(t) = \frac{1}{\varphi^3(t)} [K_0 + 2k D(t)]  \, .
\ee
\item[A3]
Determine the indefinite integral
\be \label{A-gamma}
\zeta(t) =  \!  \int \! \frac{\dif t}{t^2 \, \sqrt{K_0 + 2 k D(t)}}   \, ,
\ee
and obtain the metric function
\begin{equation}
\alpha(t,r) = t[\zeta(t) + Q(r)] \, .
\end{equation}
\end{itemize}
Then, the metric functions $\{\varphi(t), \alpha(t,r), v(t)\}$ define a T-model (\ref{metric-T-2}) that is a solution of the field equation (\ref{eq-T-2}). 

Note that the steps 1 and 2 provide a particular homogeneous KCKS T-model, and step 3 completes the nonhomogeneous solution, for which two indefinite integrals are necessary. 

In next section we will determine the general solution for $k=0$ making use of this algorithm. The case $k\not=0$ requires further analysis in order to obtain the general solution without needing any integral (see Sec. \ref{sec-solucio-knot=0} below).  


\section{The general solution for $k=0$}
\label{sec-solucio-k=0}


\subsection{Metric and hydrodynamic quantities}
\label{metric-k=0$}

The explicit general solution for the plane symmetry can be obtained by using both the Herlt algorithm and the modified Herlt algorithm. The latter provides a more direct reasoning. Indeed, note that  when $k=0$ (\ref{A-v}) and (\ref{A-gamma}) imply, respectively, $v = K_0 \varphi^{-3}$ and $\zeta = - (\sqrt{K_0}\, t)^{-1}$. Then the arbitrary functions $\varphi(t)$ and $Q(r)$, and the spatial coordinates $\{r, x ,y\}$, can be redefined by a factor in such a way that the metric line element (\ref{metric-T-2}) becomes
\begin{equation} \label{metric-T-k=0}
ds^2= -\varphi^3(t)dt^2 +  \varphi(t)([t\,  Q(r)+1]^2  dr^2 + dx^2+dy^2) \, .
\end{equation}
The unit velocity of the fluid $u = \varphi^{-3/2} \, \partial_t$ has an expansion given by
\begin{equation} \label{expansion-T-k=0}
\hspace{-0mm} \theta = \frac{1}{\varphi^{3/2}} \left(\frac32 \frac{\dot{\varphi}}{\varphi} \!+\! \frac{Q}{t\,Q+1}\right)\!  = \! \frac{1}{\varphi^{3/2}} \partial_t [\ln(\varphi^{3/2} (t\,Q+1)] \,  .
\end{equation}
And the {\em pressure} $p$ and the {\em energy density} $\rho$ are then given by 
\begin{eqnarray} \label{pressure-T-k=0}
p =  \frac{1}{\varphi^{3}} \left[\frac74 \frac{\dot{\varphi}^2}{\varphi^2} - \frac{\ddot{\varphi}}{\varphi}\right]    \,  ,  \\[1mm]
\rho =  \frac{1}{\varphi^{3}} \left[\frac34  \frac{\dot{\varphi}^2}{\varphi^2} +  \frac{\dot{\varphi}\, Q}{\varphi(t\,Q+1)}\right]    \,  .
\label{density-T-k=0}
\end{eqnarray}

On the other hand, we can specify the indicatrix function (\ref{chi-Tmodels}) in this case by calculating the implicit functions of $p$ given in (\ref{ABCcal}) in terms of $\varphi(t)$ and its derivatives: 
\begin{subequations} \label{ABCcal-02}
\begin{align}
{\cal A}(p) \equiv  \frac{4 \varphi^{10}}{\dot{\varphi}(35 \dot{\varphi}^3 + 4 \varphi^2 \dddot{\varphi}- 30\varphi\dot{\varphi} \ddot{\varphi})}  , \\[1mm] 
{\cal B}(p) \equiv {\cal A}(p)\left[p +  \frac{3 \dot{\varphi}^2}{4 \varphi^5}\right]  ,  \quad 
{\cal C}(p) \equiv {\cal A}(p) \, p \,  \frac{3 \dot{\varphi}^2}{4 \varphi^5}    .
\end{align} 
\end{subequations}

It is worth remarking that we can recover previously known T-models with plane symmetry by giving specific expressions of the function $\varphi(t)$:
\begin{itemize}
\item[(i)]
If we take $\varphi(t) = t^{-4/3}$ we obtain the dust solution. This T-model was considered by Vajk and Eltgroth \cite{Vajk} for the homogeneous case ($Q=constant$). The proper time of the fluid is $\tau = -1/t$.
\item[(ii)]
If we take $\varphi(t) = t^m$, $m >-4/3$, $m \not=0 $, we obtain the ideal T-models for $\gamma = \frac{4}{3m}+2  \not=2$ studied in \cite{FM-Tmodels} and quoted in Sec. \ref{subsec-ideal}. The proper time of the fluid is $\tau = \frac{1}{\bar{m}} t^{\bar{m}}, \ \bar{m} \equiv 1 + \frac{3m}{2}$.
\item[(iii)]
If we take $\varphi(t) = e^{2t/3}$ we obtain the ideal T-models with $\gamma =2$ studied in \cite{FM-Tmodels} and quoted in Sec.\ref{subsec-ideal}. The proper time of the fluid is $\tau = e^t$. 
\end{itemize}
\begin{figure*}
\centerline{
\parbox[c]{0.51\textwidth}{\includegraphics[width=0.50\textwidth]{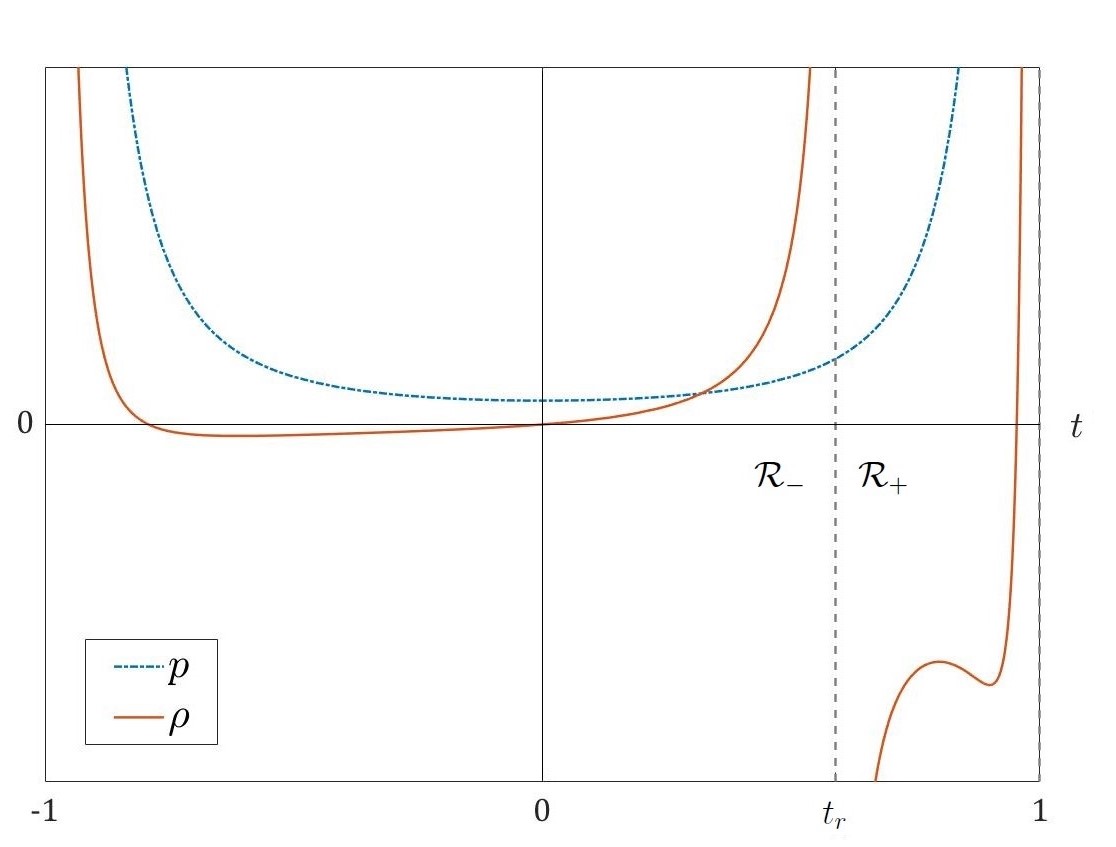}}
\parbox[c]{0.53\textwidth}{\includegraphics[width=0.495\textwidth]{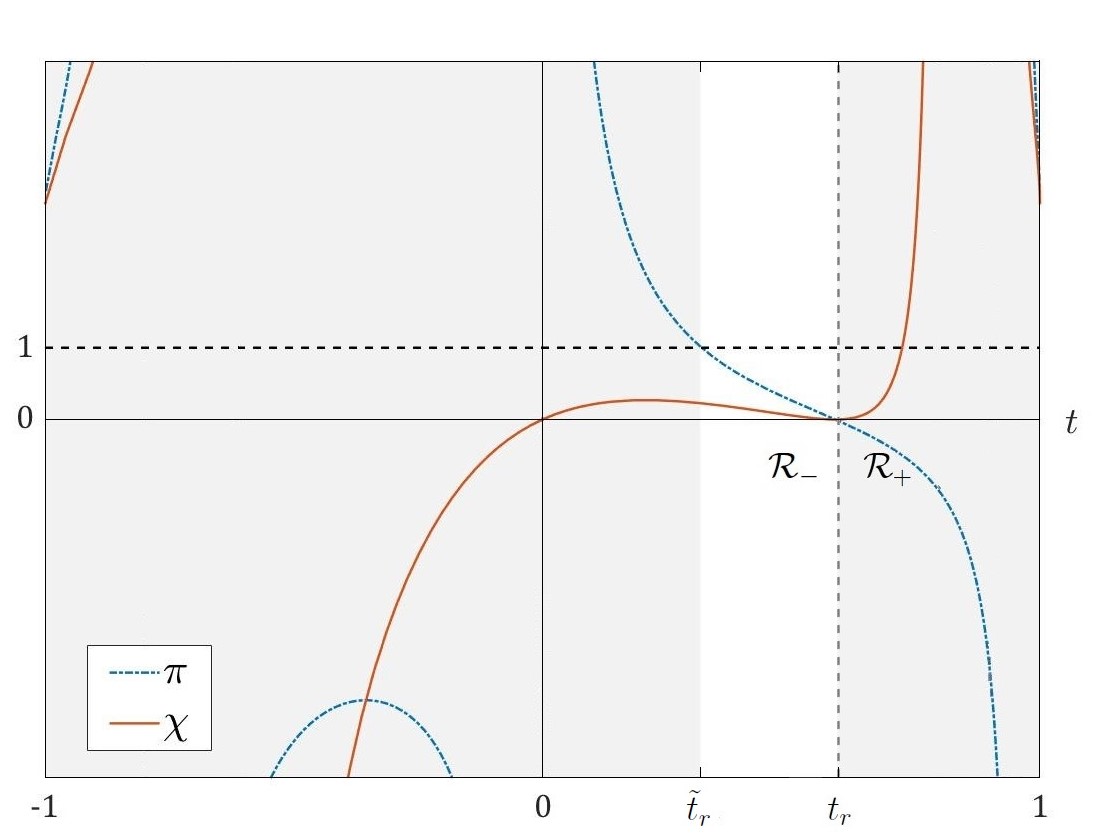}\vspace{1mm}}}
\caption{This figure shows the behavior of the hydrodynamic variables of the T-model with plane symmetry defined by the function (\ref{phi-k=0new}). We have considered the case $Q_r <0$. The case $Q_r>0$ follows by changing $t$ by $-t$. On the left, we have plotted the dependence on time of the energy density $\rho$ and the pressure $p$ for $Q_r = -1.6$. Note that $\rho$ is positive in subregion $]0, t_r[$ of the region ${\cal R}_-$. On the right, we have plotted the quotient $\pi=p/\rho$, and the square of the speed of sound  $\chi = u(p)/u(\rho)$. Notice that the energy conditions ($0 < \pi < 1$), and the causal sonic condition ($0 < \chi < 1$) only hold in the subregion $]\tilde{t}_r, t_r[$ of the region ${\cal R}_-$, where $\tilde{t}_r$ is defined by the condition $\pi(\tilde{t}_r)=1$ (unshaded interval).}
\label{Fig-2}
\end{figure*}
%


\subsection{A new solution with $k=0$}
\label{subsec-solutions-k=0$}

As an example to see how the above method to obtain the general solution for $k=0$ works, we now obtain a new solution. We take
\be \label{phi-k=0new}
\varphi(T) = \cos^{3/2}(\kappa T) \, .
\ee
Then, we obtain a T-model with $k=0$ if we replace this expression of $\varphi(T)$ in the metric line element (\ref{metric-T-k=0}). Moreover, we can analyze the physical behavior of the solution taking into account the expressions (\ref{pressure-T-k=0}) and (\ref{density-T-k=0}) of the pressure and energy density, and the expression (\ref{chi-Tmodels}, \ref{ABCcal-02}) of the indicatrix function $\chi(\rho, p)$. Nevertheless, in this case we can easily obtain the proper time $\tau$ of the fluid. Indeed, we have $d \tau = \varphi^{3/2} d T = \cos(\kappa T) d T$ and, consequently, $\tau =  \frac{1}{\kappa} \sin(\kappa T)$. Then, if we introduce the time
\be
t = \kappa \tau =  \sin(\kappa T) \in ]-1, 1[ \, ,
\ee
and we make use of the general expressions (\ref{pressure-T-1}, \ref{density-T-1}) for the hydrodynamic variables, we obtain
\begin{eqnarray} \label{pressure-T-k=0new}
p =  \frac{\kappa^2(2+ t^2)}{3(1-t^2)}     \,  ,  \\[1mm]
\rho=\frac{\kappa^2t^2}{3(1-t^2)^2}\left[1-\frac{\sqrt{1-t^2}\,Q}{t(Q\,\textrm{arcsin}\,t+\kappa)}\right] \, .
\label{density-T-k=0}
\end{eqnarray}
Note that this solution has a positive pressure everywhere, and the metric has a curvature singularity at $t=t_r\equiv\sin[-\kappa/Q(r)]$, which disconnects two spacetime regions ${\cal R}_-$  ($t<t_r$) and ${\cal R}_+$ ($t>t_r$). In the spacetime domain $Q(r)<0$ (respectively, $Q(r)>0$) the energy density is positive in the subregion $t\in\,]\,0,t_r[$ of ${\cal R}_-$ (respectively, $t\in\,]\,t_r,0[$ of ${\cal R}_+$), as the left diagram in Fig. \ref{Fig-2} shows. Moreover, there is always a spacetime domain in which the macroscopic conditions for physical reality hols if, and only if, $|\frac{Q}{\kappa}|>\frac{2}{\pi}$ (see right diagram in Fig. \ref{Fig-2}).


\section{The general solution for $k \not=0$}
\label{sec-solucio-knot=0}


\subsection{The field equation in the variables $(\varphi, \alpha, \beta)$}
\label{subsec-beta}

Now we introduce a new function $\beta(t)$ as unknown metric function. Let's define 
\begin{equation} \label{beta}
\beta(t) = v(t) \, \varphi^3(t)  > 0 \, .
\end{equation}
Then, the field equation (\ref{eq-T-2}) becomes
\be \label{eq-T-beta}
2 \beta  \, \ddot{\alpha} + \dot{\beta}\, \dot{\alpha} - 2k\, \alpha \, \varphi^2 = 0 \, .
\ee
The solution $\alpha(t)$ to this equation is of the form (\ref{alpha-1-2}), where $Q(r)$ is an arbitrary real function, and $\alpha_i(t)$ being two particular solutions to the Eq. (\ref{eq-T-beta}). A straightforward calculation shows that, if $\alpha_1(t)$ fulfills (\ref{eq-T-beta}), then another independent solution can be written as $\alpha_2(t) = \gamma(t) \alpha_1(t)$ where $\gamma(t)$ meets the equation
\be \label{eq-T-gamma}
\dot{\gamma}^2 \, \alpha_1^4 \, \beta = 1   \, .
\ee

It is worth remarking that (\ref{eq-T-beta}) is an algebraic equation for the function $\varphi(t)$. Consequently, $\varphi(t)$ can be obtained without quadratures in terms of $\beta(t)$ and $\alpha_1(t)$. This fact and the Eq. (\ref{eq-T-gamma}) allow us to obtain the general solution for $k \not=0$ without needing to calculate any integral. Hereunder we develop two algorithms that determine this solution in terms of an arbitrary function of time.


\subsection{The $\gamma$-algorithm}
\label{subsec-gamma-al}

Note that any solution $\alpha(t)$ to Eq. (\ref{eq-T-beta}) is a nonconstant function when $k \not= 0$. Thus, we can take the time coordinate $t$ such that
\be \label{alpha=t}
\alpha_1(t) = t   \, .
\ee
Then, equations (\ref{eq-T-beta}) and (\ref{eq-T-gamma}) become, respectively,
\be \label{eq-T-beta-gamma}
\dot{\beta} = 2k \, t \, \varphi^2 \, , \qquad   \dot{\gamma}^2 \, t^4 \, \beta = 1   \, .
\ee
From these expressions we can perform the following algorithm to obtain the general solution of the field equations:
\begin{itemize}
\item[G1]
Choose two arbitrary real functions $\{\gamma(t), Q(r)\}$. 
\item[G2]
Determine the function 
\be
\beta(t) =  \frac{1}{t^4 \, \dot{\gamma}^2(t)}  \, .
\ee
\item[G3]
Determine the metric functions 
\begin{subequations}
\begin{align}
v(t) = \frac{\beta(t)}{\varphi^3(t)} \, , \qquad 
\varphi(t) =  \sqrt{\frac{|\dot{\beta}(t)|}{2 t}}  \, , \\[1mm]
\alpha(t) = t \,[1 + \gamma(t) Q(r)]\, .  \qquad \qquad 
\end{align}
\end{subequations}
\end{itemize}
Then, the triad $\{\varphi(t), \alpha(t,r), v(t)\}$ defines a T-model (\ref{metric-T-2}) which is a solution of the field equation (\ref{eq-T-2}) for
\begin{itemize}
\item[-]
spherical symmetry, $k \! = \! +1$, in the spacetime domain where $\dot{\beta}(t) > 0$, 
\item[-]
hyperbolic symmetry, $k \!=\! -1$, in the spacetime domain where $\dot{\beta}(t) < 0$. 
\end{itemize}
Moreover, if $\dot{\beta}(t_1) =0$ the metric is singular at $t_1$.

We can recover previously known T-models with nonplane symmetry by giving specific expressions of the function $\gamma(t)$:
\begin{itemize}
\item[(i)]
If we take $\gamma(t) = t^{-\frac{2n}{n-1}}$ we obtain the McVittie-Wilshire-Herlt solution quoted in Sec. \ref{subsec-Herlt}. The proper time of the fluid is $\tau = \frac{[k(n-1)]^{3/4}}{\sqrt{2n}}t^{\frac{1}{n-1}}$.
\item[(ii)]
If we take $\gamma(t) \!= \! \arcsin \! \sqrt{1\!-\!\frac{1}{\kappa^2 t^2}}$ we obtain the sphe\-rically symmetric model (\ref{phi-knot0}, \ref{omega-k=1}) obtained in Sec. \ref{subsec-v=1-knot=0}. The proper time of the fluid is $\tau = \sqrt{\kappa^2 \! - \! \frac{1}{t^2}}$.  
\item[(iii)]
If we take $\gamma(t) \!= \! \arsinh \! \sqrt{\frac{1}{\kappa^2 t^2}\!-\!\ 1}$ we obtain the hyperbolically symmetric model (\ref{phi-knot0}, \ref{omega-k=-1+}) obtained in Sec. \ref{subsec-v=1-knot=0}. The proper time of the fluid is $\tau = \sqrt{\frac{1}{t^2} \! - \! \kappa^2} $.  
\item[(iv)]
If we take $\gamma(t) \!= \! \arcosh \! \sqrt{\frac{1}{\kappa^2 t^2}\!+\!\ 1}$ we obtain the hyperbolically symmetric model (\ref{phi-knot0}, \ref{omega-k=-1-}) obtained in Sec. \ref{subsec-v=1-knot=0}. The proper time of the fluid is $\tau = \sqrt{\frac{1}{t^2} \! + \! \kappa^2} $.  
\end{itemize}
%


\subsection{The $\xi$-algorithm}
\label{subsec-xi-al}

If $\alpha_i(t)$ are two independent solutions to the  equation (\ref{eq-T-beta}), then $\alpha_2 (t)= \alpha_1(t) \gamma(t)$, with $\dot{\gamma} \not=0$. Thus, we can take the time coordinate $t$ such that
\be \label{alpha=t}
\gamma(t) = t   \, .
\ee
Then, if $\xi (t)= 1/\alpha_1(t)$, equations (\ref{eq-T-beta}) and (\ref{eq-T-gamma}) become, respectively,
%
%
\be \label{eq-T-beta-xi}
 \beta = \xi^4 \, , \qquad \xi^3 \ddot{\xi} +  k \varphi^2 = 0   \, .
\ee
From these expressions we can perform the following algorithm to obtain the general solution of the field equations:
\begin{itemize}
\item[X1]
Choose two arbitrary real functions $\{\xi(t), Q(r)\}$. 
\item[X2]
%
%
Determine the metric functions 
\begin{subequations}
\begin{align}
v(t) \!=\! \frac{1}{\sqrt{\xi(t)\, |\ddot{\xi}(t)|^3}}  , \quad \varphi(t)\! = \! \sqrt{\xi^3(t)\, |\ddot{\xi}(t)|}   , \\[1mm]
\alpha(t) = \frac{1 + t\,  Q(r)}{\xi(t)}\, .  \qquad \qquad  
\end{align}
\end{subequations}
\end{itemize}
Then, the triad $\{\varphi(t), \alpha(t,r), v(t)\}$ defines a T-model (\ref{metric-T-2}) which is a solution of the field equation (\ref{eq-T-2}) for
\begin{itemize}
\item[-]
spherical symmetry, $k \! = \! +1$, in the spacetime domain where $\ddot{\xi}(t) < 0$, 
\item[-]
hyperbolic symmetry, $k \!=\! -1$, in the spacetime domain where $\ddot{\xi}(t) > 0$. 
\end{itemize}
Moreover, if $\ddot{\xi}(t_1) =0$ the metric is singular at $t_1$.

\begin{figure*}
\centerline{
\parbox[c]{0.51\textwidth}{\includegraphics[width=0.49\textwidth]{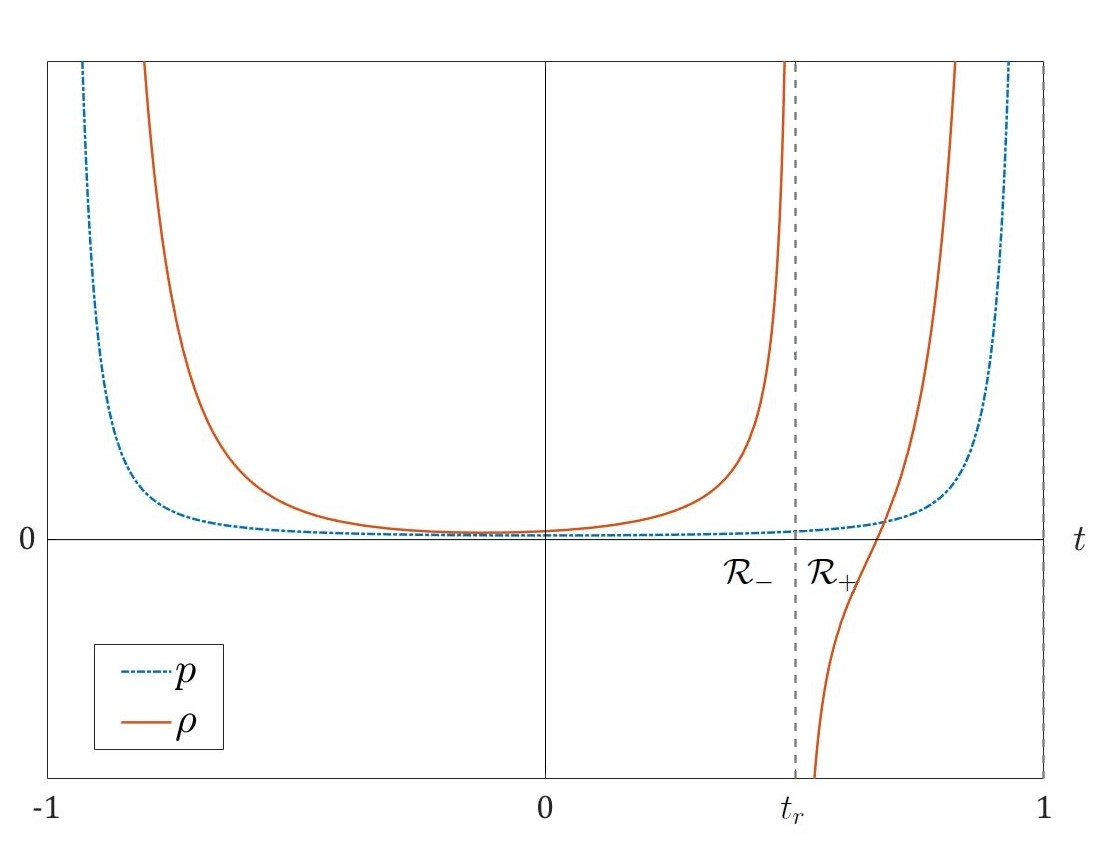}}
\parbox[c]{0.53\textwidth}{\includegraphics[width=0.495\textwidth]{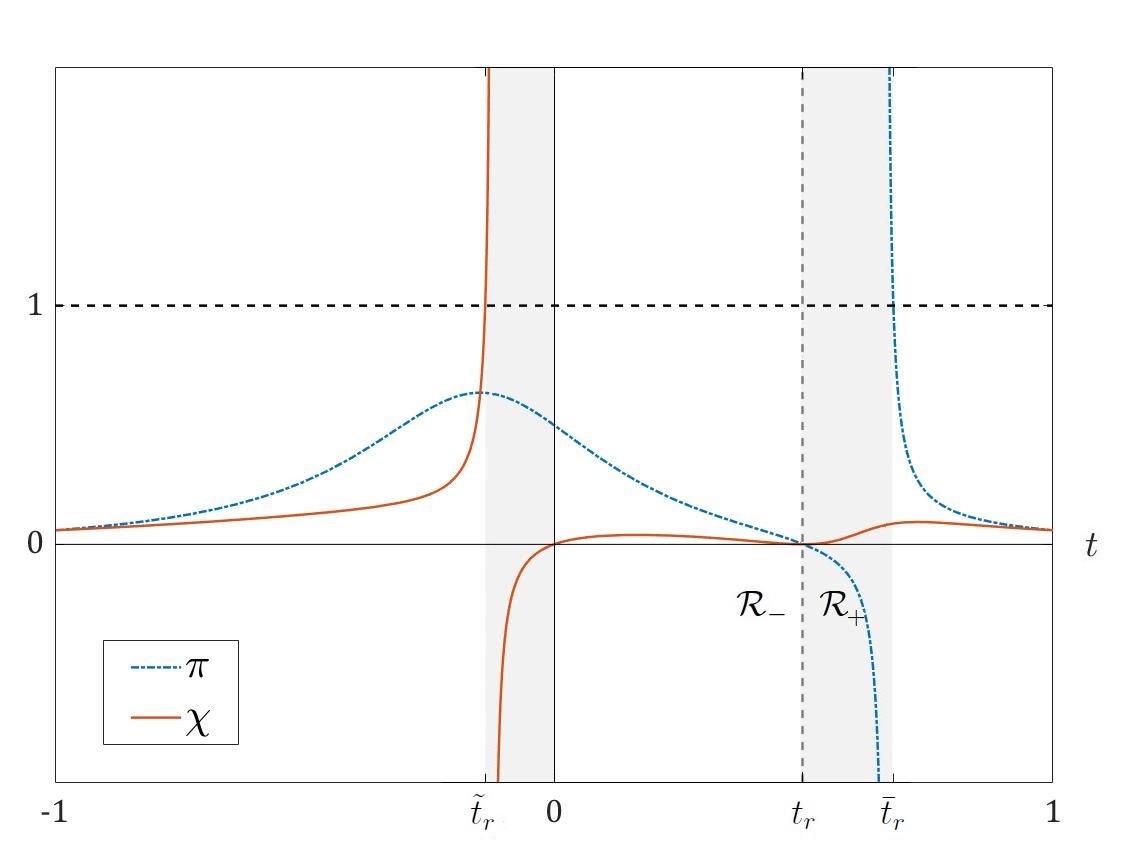}}}
\caption{This figure shows the behavior of the hydrodynamic variables of the T-model with spherical symmetry determined by the $\xi$-algorithm with $\xi = 1-t^2$. We have considered the case $Q_r <0$. The case $Q_r<0$ follows by changing $t$ by $-t$. On the left, we have plotted the dependence on time of the energy density $\rho$ and the pressure $p$ for $Q_r = -2$. Note that $\rho$ is positive in the whole region ${\cal R}_-$ and in a part of the region ${\cal R}_+$. On the right, we have plotted the quotient $\pi=p/\rho$, and the square of the speed of sound  $\chi = u(p)/u(\rho)$. Notice that the energy conditions ($0 < \pi < 1$), and the causal sonic condition ($0 < \chi < 1$) only hold in the subregions $]-1,\tilde{t}_r[$ and $]\,0,t_r[$ of ${\cal R}_-$, and $]\,\bar{t}_r,1[$ of ${\cal R}_+$ (unshaded intervals), where $\tilde{t}_r$ and $\bar{t}_r$ are defined by the conditions $\chi(\tilde{t}_r)=1$ and $\pi(\bar{t}_r)=1$, respectively.}
\label{Fig-3}
\end{figure*}

We can recover previously known T-models with nonplane symmetry by giving specific expressions of the function $\xi(t)$:
\begin{itemize}
\item[(i)]
If we take $\xi(t) = t^{\frac{n-1}{2n}}$ we obtain the McVittie-Wilshire-Herlt solution quoted in Sec. \ref{subsec-Herlt}, The proper time of the fluid is $\tau = -\frac{[k(n-1)]^{3/4}}{\sqrt{2n}}t^{-\frac{1}{2n}}$.

\item[(ii)]
If we take $\xi(t) = \kappa \cos t$ we obtain the spherically symmetric model (\ref{phi-knot0}, \ref{omega-k=1}) obtained in Sec. \ref{subsec-v=1-knot=0}. The proper time of the fluid is $\tau = \kappa \sin t$.  
\item[(iii)]
If we take $\xi(t) = \kappa \cosh t$ we obtain the hyperbolically symmetric model (\ref{phi-knot0}, \ref{omega-k=-1+}) obtained in Sec. \ref{subsec-v=1-knot=0}. The proper time of the fluid is $\tau = \kappa \sinh t$.
\item[(iv)]
If we take $\xi(t) = \kappa \sinh t$ we obtain the hyperbolically symmetric model (\ref{phi-knot0}, \ref{omega-k=-1-}) obtained in Sec. \ref{subsec-v=1-knot=0}. The proper time of the fluid is $\tau = \kappa \cosh t$.  
\end{itemize}
%


\subsection{A new spherically symmetric solution}
\label{subsec-new-knot=0}

Now we consider an example to see how the above algorithms to obtain the general solution for $k\not=0$ work. We take
\be \label{xi-knot0new}
\xi(t) = 1- t^n \, .
\ee
Then, we get a T-model with $k\not=0$ if we apply the $\xi$-algorithm. We have that for any $n$ out of the range $[0,1]$ the solution is spherically symmetric ($\ddot{\xi}(t)<0$) in a spacetime domain. Moreover, we can analyze the physical behavior of the solutions taking into account the general expressions (\ref{pressure-T-1}) for the pressure and (\ref{density-T-2}) for the energy density, and the expression (\ref{chi-Tmodels}, \ref{ABCcal}) of the indicatrix function $\chi(\rho, p)$. The metric has a curvature singularity at $t=t_r\equiv -1/Q(r)$, which disconnects two spacetime regions ${\cal R}_-$  ($t<t_r$) and ${\cal R}_+$ ($t>t_r$). 

For sake of simplicity, we now focus on the case $n=2$. Then, the solution is spherically symmetric in the interval $t\in\,]-1,1[$, and the pressure and the energy density take the expressions
\begin{eqnarray} \label{pressure-T-knot=0new}
p =  \frac{20-17 t^2}{8 \sqrt{2}(1-t^2)^{5/2}}      ,  \qquad \\[1mm]
\rho=\frac{1}{8 \sqrt{2}(1-t^2)^{5/2}}\! \left[51 t^2 \!-\! \frac{4(1-t^2)(Q \, t-2)}{1 + Q \, t}\right] \! . \qquad
\label{density-T-k=0}
\end{eqnarray}
Note that this solution has a positive pressure everywhere, and when $Q(r)<0$ (respectively, $Q(r)>0$) the energy density is positive in the region ${\cal R}_-$ (respectively, in ${\cal R}_+$) and in a part of the region ${\cal R}_+$ (respectively, in ${\cal R}_-$), as the left diagram in Fig. \ref{Fig-3} shows. Moreover, there is always a spacetime domain in which the macroscopic conditions for physical reality hold (see right diagram in Fig. \ref{Fig-3}).


\section{Discussion}
\label{sec-remarks}

The metric functions defining the metric line element (\ref{metric-T-1})  of a T-model are submitted to a differential equation (\ref{eq-T-1}). Herlt \cite{Herlt} proposed an integration algorithm that showed that this field equation can be solved by quadratures. Here, in Sec. \ref{sec-integracio-1}, we have revisited the Herlt approach and we have proposed a modified procedure. In both algorithms the solution is obtained by calculating two indefinite integrals.

By undertaking an in-depth study of the field equation and redefining the unknown metric functions, we have established some algorithms that solve the equation without calculating any integral. Thus, we give the explicit expression of the general solution (in Sec. \ref{sec-solucio-k=0} for plane symmetry, $k=0$, and in Sec. \ref{sec-solucio-knot=0} for $k \not=0$) depending on an arbitrary function of time and an arbitrary function $Q(r)$ of the spatial coordinate $r$. 

We have recovered some known solutions and we have obtained new ones by applying any of the above quoted algorithms. The physical meaning of these T-models can be analyzed {\it a posteriori} by using our hydrodynamic approach to the perfect fluid solutions \cite{CFS-LTE, CFS-CC, CFS-RSS}. Nevertheless, it would be appropriate to be able to impose specific physical or geometrical properties established {\it a priori}, as we have performed with the ideal T-models analyzed in \cite{FM-Tmodels} and quoted here in Sec. \ref{subsec-ideal}. This aim justifies having presented different integration methods here so that we can choose the one that is the most suitable for the restrictions we impose. 

The Herlt and our modified Herlt algorithms provide a (particular) homogeneous solution with a quadrature, and the general solution of the nonhomogeneous case follows by obtaining another indefinite integral. The general solution of the homogeneous T-models corresponds with the nonhomogeneous one for the case $Q(r) = constant$. Consequently, our study also provides the general solutions of the KCKS T-models.

The Szekeres-Szafron solutions of class II \cite{Krasinski-Plebanski, Krasinski, Szekeres, Szafron, FS-SS} are a generalization without symmetries of the T-models. A thermodynamic analysis of these solutions shows \cite{CFS-PSS}  that three subfamilies in local thermal equilibrium can be considered: the singular models, the regular models and the T-models. The latter are the object of the present paper and have been analyzed from a thermodynamical point of view in \cite{FM-Tmodels}. The Szekeres-Szafron singular and regular models have been studied in \cite{CFS-PSS} and \cite{CFS-RSS}, respectively. In both cases the metric line element and the field equation are similar to those of the T-models by changing the function $Q(r)$ by 
\be
\tilde{Q}(r,x,y) = \frac12 U(r) (x^2 + y^2) + V_1 (r) x +  V_2 (r) y + 2\,W(r) ,
\ee
where $V_1 (r) , V_2 (r), W(r) $ are arbitrary real functions, and $U(r) + k W(r) = 1$ for the regular models, and $U(r)=0$ and $k=0$ for the singular models. Consequently, all the integration algorithms obtained in this paper for the T-models also apply for the thermodynamic class II Szekeres-Szafron solutions (singular and regular models). These solutions without symmetries can be obtained from a T-model by changing $Q(r)$ by $\tilde{Q}(r,x,y)$.


\begin{acknowledgements}
This work has been supported by the Spanish Ministerio de Ciencia, Innovaci\'on y Universidades and the Fondo Europeo de Desarrollo Regional, Projects No. PID2019-109753GB-C21 and No. PID2019-109753GB-C22, the Generalitat Valenciana Project No. AICO/2020/125 and the University of Valencia Special Action Project No. UV-INVAE19-1197312. 
\end{acknowledgements}


\bibliography{PRD_Tmodels}


\end{document}